\documentclass[epsfig,rotating]{elsart} %[epsfig]

\usepackage{epsfig}
\usepackage{amsmath} % ams business is for use of 'cases'
\usepackage{amsfonts}
\usepackage{amssymb}
\usepackage{setspace}

\begin{document}
\newcommand{\pwlw}{{power-law }}
\newcommand{\tg}{{\gamma}}
\newcommand{\tgH}{{\hat{\gamma}}}
\newcommand{\otp}{{\overline{TP}}}

\begin{frontmatter}
\journal{Astroparticle Physics}



\title{Power Laws and the Cosmic Ray Energy Spectrum}


\author[New Mexico]{J.~D.~Hague\thanksref{email} }
\author[New Mexico]{B.~R.~Becker}
\author[New Mexico]{M.~S.~Gold}
\author[New Mexico]{J.A.J.~Matthews}

\address[New Mexico]{University of New Mexico, Department of Physics
and Astronomy, Albuquerque, New Mexico, USA}

\thanks[email]{Corresponding author, E-mail: \texttt{jhague@unm.edu} }

\date{\today}
%\date{June 30, 2006}

\begin{abstract}
Two separate statistical tests are applied to the AGASA and preliminary Auger Cosmic Ray Energy spectra 
in an attempt to find deviation from a pure power-law. The first test is constructed from the probability distribution 
for the maximum event of a sample drawn from a power-law. The second employs the TP-statistic, a function defined 
to deviate from zero when the sample deviates from the power-law form, regardless of the value of the power index. 
The AGASA data show no significant deviation from a power-law when subjected to both tests.
Applying these tests to the Auger spectrum suggests deviation from a power-law. However, potentially 
large systematics on the relative energy scale prevent us from drawing definite conclusions at this time.
\end{abstract}


\end{frontmatter}

%=========================================================================
\section{Introduction}
\vspace{0.1in}
Nature offers a wide range of phenomena characterized by \pwlw distributions: diameter of moon craters, intensity of solar flares, 
the wealth of the richest people\cite{Newm} and intensity of terrorist attacks\cite{Claus}, to name a few. 
These distributions are so-called 
{\it heavy-tailed}, where the fractional area under the tail of the distribution is larger than that of a gaussian and there is thus 
more chance for samples drawn from these distributions to contain large fluctuations from the mean.
Anatomical\footnote{Well known small deviations from a pure \pwlw are dubbed ``The Knee'' and ``The Ankle.''} 
defects aside, the cosmic ray (CR) energy spectrum follows a \pwlw for over 
ten orders of magnitude. The predicted abrupt deviation at the very highest energies (the GZK-cutoff\cite{G,ZK})  
has generated a fury of theoretical and experimental work in the past half century. Recently,  Bahcall\cite{Bahc} and Waxman 
(2003) have asserted that the observed spectra (except AGASA) are consistent with the 
expected flux suppression above $5 \times 10^{19}$eV. However, the incredibly low fluxes combined with 
as much as $\sim$50\% uncertainty in the absolute energy determination means that there has yet to be a complete consensus on 
the existence of the GZK-cutoff energy.

With this in mind, we consider statistics which suggest an answer to a different question: {\it Do the observed CR 
spectra follow a power-law?} Specifically, these studies are designed to inquire whether or not there is a 
flux deviation relative to the \pwlw form by seeking to minimize the influence of the underlying parameters.

The two experimental data sets considered in this study are  the AGASA\cite{AGASA} experiment
and the preliminary flux result of the Pierre Auger Observatory\cite{Auger,Yama}. The discussion in $\S$\ref{sec:data} uses these 
spectra to introduce and comment on the \pwlw form.
The first distinct statistical test is applied to this data in $\S$\ref{sec:DLV} where we explore the 
distribution of the largest value of a sample drawn from a power-law. In $\S$\ref{sec:TP} we apply the 
TP-statistic to the CR flux data. This statistic is asymptotically zero for pure \pwlw samples {\it regardless} 
of the value power index and therefore offers a (nearly) parameter free method of determining deviation from the 
\pwlw form. The final section summarizes our results.

%=========================================================================
\section{The Data}\label{sec:data}
\vspace{0.1in}

A random variable $X$ is said to follow a \pwlw distribution if the probability 
of observing a value between $x$ and $x + dx$ is $f(x)dx$ where $f(x) = C x^{-\tg}$.
Normalizing this function such that $\int_{x_{min}}^{\infty} f(x) dx = 1$ gives,
\begin{equation}
f_{X}(x) = \frac{\tg - 1}{x_{min}} \left( \frac{x}{x_{min}} \right) ^{-\tg}.   \label{eq:pwlpdf}
\end{equation}
It is convenient to choose $z = x/x_{min} \Rightarrow dz = dx/x_{min}$, $1 \leq z < \infty$ and doing so yields 
\begin{equation}
f_{Z}(z) = (\tg - 1) z^{-\tg}.   \label{eq:pwlpdfZ}
\end{equation}
For reference, one minus the cumulative distribution function $F_{Z}(z)$ is given by, 
\begin{equation}
1 - F_{Z}(z) = \int_{z}^{\infty} f_{Z}(y) dy = z^{1-\tg}. \label{eq:pwlcdf}
\end{equation}

Taking the log of both sides of equation (\ref{eq:pwlpdf}) yields 
\begin{equation}
\log f(x) = \log A - \tg \log x, \label{eq:loglogfit}
\end{equation}
where $A$ is an overall normalization parameter, and suggests a method of estimating   
$\tg$; the {\it power index} is the slope of the best fit line to the logarithmically binned data 
(i.e. bin-centers with equally spaced logarithms). In what follows, we refer to the logarithmically binned estimate\cite{maxlik} 
of the power index as $\tgH$ and assume that the typical $\chi^{2}$/NDF is indicative of the goodness of fit. The 
fitting is done with two free parameters, namely $A$ and $\tg$.

\begin{figure}[h]
\begin{center}
\includegraphics[width=11.1cm, height=7cm]{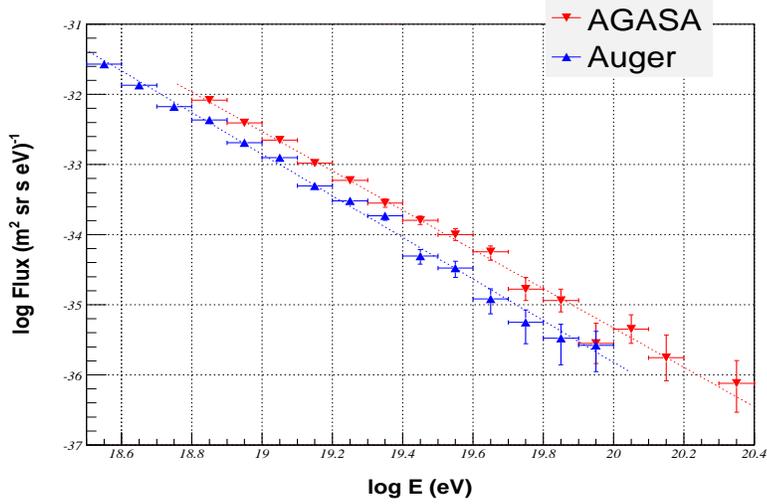} %
\caption{\label{fig:specs} This figure displays published AGASA\cite{AGASA} and Auger\cite{Auger} CR energy spectra. 
Both axis have logarithmic scales to illustrate the \pwlw behavior. 
The vertical axis is the flux $J$ in (m$^2$ sr sec eV)$^{-1}$ and the horizontal axis is the energy in eV.
The best fit lines (see \ref{eq:loglogfit}) have slope $\hat{\tg}_{AGASA} = 2.80 \pm 0.23$ and 
$\hat{\tg}_{Auger} = 2.97 \pm 0.12$ (statistical error only).}
\end{center}
\end{figure}

The energy flux of two publicly available data sets are shown in Fig. \ref{fig:specs}. The 
the red point-down triangles represent the log$_{10}$ of the binned AGASA flux values in units of 
(m$^2$ sr sec eV)$^{-1}$ and the blue point-up triangles correspond to the Auger flux. 
The vertical error bars on each bin reflect the Poisson error based on the number of events in that bin. 
The log-binned estimates for each complete CR data set are the slopes of the dashed lines plotted in Fig. \ref{fig:specs}. 

\begin{figure}[h]
\begin{center}
\includegraphics[width=11.1cm, height=7cm]{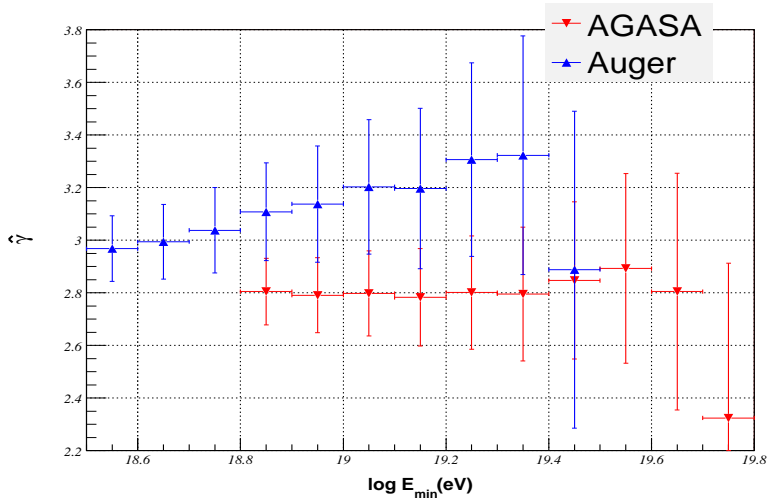} %
\caption{\label{fig:Gests} To check the stability of $\tgH$ we estimate the power index as a function of the minimum 
energy $E_{min}$ considered for the AGASA and Auger CR data sets; see Fig.\ref{fig:specs}. The left most point is the slope of the 
best fit lines plotted in  Fig.\ref{fig:specs}. 
The vertical error bars represent $1\sigma_{\tgH}$ deviation.}
\end{center}
\end{figure}

In order to check the stability of to bound on our estimate, we compute the estimated power index 
$\tgH$ as a function of the minimum 
energy $E_{min}$ considered for each of the two CR data sets. The left-most blue (red) point in Fig. \ref{fig:Gests} 
shows $\tgH$ for the Auger (AGASA) data taking into account all of the bin values above 
$\log E_{min} = 18.5$ ($\log E_{min} = 18.8$), the next point to the right represents that for all bins above 
$\log E_{min} = 18.6$ ($\log E_{min} = 18.9$), and so on. The vertical error bars on these points represent the $1\sigma_{\tgH}$ 
error of the estimate. To ensure an acceptable chi-squared statistic, we demand that at least five bins be considered, thereby truncating 
$E_{min}$ at $\log E_{min} = 19.4$ for the Auger and $\log E_{min} = 19.7$ for the AGASA data set. The $\chi^{2}$/NDF for the 
left-most points is $\sim0.3$ and it increases to $\sim2.5$ for the right-most for both experiments. 
We note that these estimates do not vary widely for the lowest $E_{min}$'s and that the values of $\tgH$ from these experiments are 
consistent.

The analyses discussed in $\S$\ref{sec:DLV} and $\S$\ref{sec:TP} will depend on the total number of events in the 
data set. Since these numbers are not published we use a simple method for estimating them from the CR flux data. 
If the exposure is a constant function of the energy, then we may 
take the flux $J$ to be proportional to the number of events in the bin and the exposure $\eta$, namely 
$N = J \eta E_{bin-center} \ln(10)/10$. The Auger exposure is reported to be constant over the energy range 
reported with $\eta_{Auger} = 5.5 \times 10^{16}$ (m$^2$ sr sec). The AGASA collaboration report flux data 
all the way down to $\log E_{min} = 18.5$ but the exposure of the experiment can be considered approximately constant 
only for energies above  $\log E_{min} = 18.8$ (see Fig. 14 of \cite{AGASA}) where $\eta_{AGASA} = 5.1 \times 10^{16}$ (m$^2$ sr sec).
Using this method we get a total of 3567 events with $E \geq 10^{18.5}$ for the Auger flux and 1914 with $E \geq 10^{18.8}$ for the 
AGASA experiment.

%=========================================================================
%=========================================================================
\section{ The Distribution of the Largest Value }\label{sec:DLV}
\vspace{0.05in}

As evidence suggestive of a GZK-cutoff, an often cited quantity is the flux suppression, or the ratio of the flux one 
would expect from a power-law to that actually observed above a given maximum, say, $z_{max}$. Since $J \propto N$ one 
may estimate the flux suppression by estimating the number of events $N_{sup}$ out of $N_{tot}$ expected above a given 
maximum as $N_{sup} = N_{tot}[1-F_{Z}(z_{max})] = N_{tot}z_{max}^{1-\gamma}$. Thus, the bin with minimum $=z_{max}$ and 
maximum $\rightarrow \infty$ would have a height $= N_{sup}$ if the data continued to follow a \pwlw above $z_{max}$. 
As a test statistic for this quantity, one may consider the Poissonian probability that the bin height could statistically 
fluctuate to zero, namely $\mathcal{P}(0, N_{sup}) = $ exp$[-N_{tot}z_{max}^{1-\gamma}]$. 

In this section we derive a similar test statistic based on the distribution of the maximum event from a \pwlw sample. 
The statistic discussed here approaches $\mathcal{P}(0, N_{sup})$ for large $N_{tot}$ and 
allows us to show that the estimation errors associated with $\tgH$ are enough to disallow any significant 
conclusion about the presence of flux suppression for the highest energy CR's.

\begin{figure}[h]
\begin{center}
\includegraphics[width=9.5cm, height=6cm]{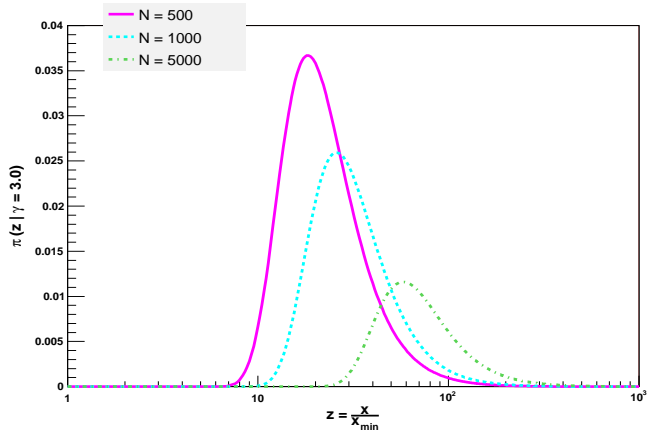}
\caption{\label{fig:maxPDF} A plot of the probability distribution of the maximum of a sample drawn from a \pwlw  
with power index $\tg = 3.0$. This is the distribution $\pi(z)$ defined in equation (\ref{eq:maxesZ}) where $z$ is the ratio 
of the maximum to the minimum. 
The sample sizes are $N=500,\,1000$ and $5000$.}
\end{center}
\end{figure}

The form of the \pwlw distribution allows us to calculate the pdf of the largest value, $X_{max}$, out of N events.
Using the equations (\ref{eq:pwlpdf}) and (\ref{eq:pwlcdf}) we can say that the probability that any one value falls 
between $x$ and $x+dx$ {\it and} that all of the others are less than it is $f(x)dx \times F(x)^{N-1}$. There are N ways
to choose this event and so the probability for the largest value to be between $x$ and $x+dx$ is
\begin{equation}
\pi(x)dx = N f(x) F(x)^{N-1}dx. \nonumber %\label{eq:maxes} 
\end{equation}
In terms of the ratio $z$, this can be written as
\begin{equation}
\pi(z)dz = N (\tg - 1) z^{-\tg} \left( 1 - z^{1-\tg} \right)^{N-1}dz. \label{eq:maxesZ}
\end{equation}
Fig. \ref{fig:maxPDF} contains a plot 
of this distribution for $\tg = 3.0$ with three choices of N. The glaring implication of this plot is that even 
for ``small'' N nearly all of the integral of $\pi(z)$ is above $z \sim 10$. This implies that the probability of 
the maximum energy event falling below 10 times the minimum is very small, for a \pwlw with these parameters.

Motivated by the location and shape of $\pi(z)$ we consider 
the probability $P$ that the maximum ratio from a given sample $Z_{max}$ is less than or equal to a particular value
\footnote{For large $N_{tot}$, equation (\ref{eq:Pxc}) approaches the Poisson probability mentioned above; 
$\left[ 1 - z_{max}^{1-\tg} \right]^{N_{tot}} \rightarrow \exp[-N_{tot}z_{max}^{1-\tg}] = \mathcal{P}(0, N_{sup})$.}, 
say $z$, 
in a convenient form as 
\begin{equation}
P(Z_{max} \leq z) = \int_{1}^z \pi(t) dt \, = \, \left[ 1 - z^{1-\tg} \right]^{N}. \label{eq:Pxc}
\end{equation}
Indeed, with $\tg =3.0$ (as in Fig. \ref{fig:maxPDF}),  $P(Z_{max} \leq 10) = 6.6\times10^{-3}$ for $N=500$,
$4.3\times10^{-5}$ for $N=1000$ and is $1.4\times10^{-23}$ for $N=5000$.
Another way to say this is that if one were to generate $10^{5}$ {\it sets} of events, each containing 1000 events
drawn from a pure \pwlw with $\tg = 3.0$, $\sim99.99\%$ of these sets would have a maximum element 
with a value greater than 10 times the minimum. For 500 events/set the fraction decreases to $\sim99.34\%$. 
Such simulations were carried out in preparation for this note and the results were consistent with equation (\ref{eq:Pxc}). 

To apply this idea to the CR spectrum we consider the following null hypothesis: {\it The flux of CR's follow 
a \pwlw with index $\tgH$ for all energies greater than a given minimum.}
As a test statistic for this hypothesis we use $P$, as defined in equation (\ref{eq:Pxc}), with the interpretation that
if the null hypothesis is true then $P$ is the probability that the ratio of the maximum energy to the minimum is 
less than or equal to the observed ratio. 
Typically, the null hypothesis is rejected at the 5\% significance level (S.L.) 
if $P\leq0.05$.

To calculate the value of $P$ for the observed data sets we need three pieces of information:  
the ratio of the maximum observed value to the minimum $z^{obs}_{max}$, 
the number of events $N_{tot}$ with values in the interval $[1,z_{max}]$ and
a reasonable guess for the power index $\tg$.
Since a larger $z_{max}$ will lead to a larger value of $P$ we will conservatively 
take the highest energy AGASA (resp. Auger) event to fall on the upper edge of the highest energy bin. 
The method of determining the number of events in each bin is described in $\S$\ref{sec:data} and here the parameter 
$N_{tot}$ represents the total number above a given minimum. We will use the logarithmically binned estimates and 
errors of $\tgH$ discussed in $\S$\ref{sec:data}.

\begin{figure}[h]
\begin{center}
\includegraphics[width=11.1cm, height=7cm]{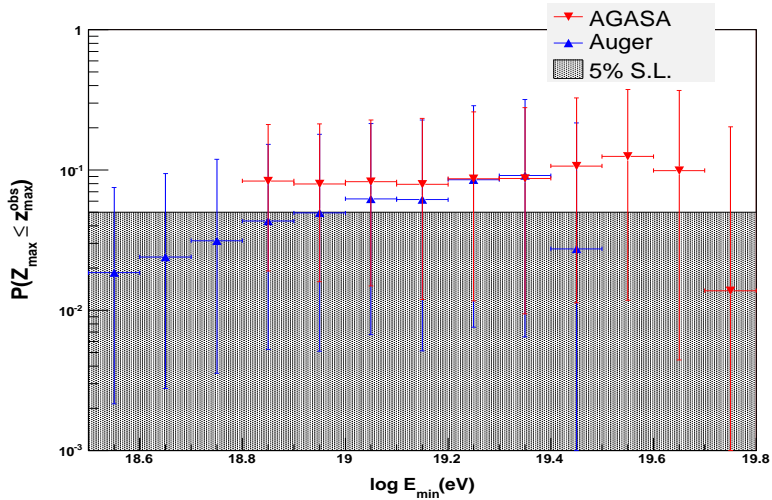} 
\caption{\label{fig:PmaxLast} A plot of the probability that the maximum of a sample drawn from a \pwlw 
will be less than or equal to the maximum observed by the Auger ($E_{max} = 10^{20}$, blue point-up) 
and AGASA ($E_{max} = 10^{20.4}$, red point-down) experiments
as a function of the minimum energy considered. The vertical error bars represent the effect of a 
$1\sigma_{\tgH}$ deviation and the hatched area shows the 5\% significance level.}
\end{center}
\end{figure}

The plot in Fig. \ref{fig:PmaxLast} shows 
$P(Z_{max} \leq z^{obs}_{max})$ given $N_{tot}$ and $\tgH$
as a function of minimum energy considered for each of the CR data sets in Fig. \ref{fig:Gests}. 
In particular, for each $E_{min}$ the values 
of $N_{tot}$, $z^{obs}_{max}$ and $\tgH \pm \sigma_{\tgH}$ are estimated 
from the CR flux and the resulting $P$ are plotted for the Auger (blue) and AGASA (red) data. 
For example, the left-most Auger point represents the probability that if $N_{tot} = 3567$ events are drawn from a 
\pwlw with $\tgH = 2.97^{+0.12}_{-0.12}$ then there is a $1.9^{+5.7}_{-1.6}\%$ chance that the maximum log-ratio $\log z_{max}$ 
would be less than or equal to that reported by the Auger experiment, namely $\log z^{obs}_{max} = \log 10^{20}/10^{18.5}$(eV).
Taken at face value, one may reject the null hypothesis at the 5\% S.L. for this data set. The left-most AGASA point represents 
the same probability for the complete set of AGASA data, namely $P(\log Z_{max} \leq \log 1.6) = 8.4^{+13}_{-6.5}\%$ for 
$N_{tot} = 1914$ events drawn from a \pwlw with $\tgH = 2.80^{+0.13}_{-0.13}$. Thus we cannot reject the null hypothesis 
for the AGASA data.

The upper (lower) vertical error bars  depicted in Fig. \ref{fig:PmaxLast} 
represent the value 
of $P$ if we have under (over) estimated the power index by $\sigma_{\tgH}$, that is if $\tg = \tgH \pm \sigma_{\tgH}$, 
keeping the log-ratio and the total number of events constant. (The possible errors in the total number of events are 
on the order of a few percent and are negligible.)  Since the fitting 
scheme considers successively lower energy bins, the points (and errors) 
for each experiment plotted in Fig. \ref{fig:PmaxLast} are highly correlated.  
The upper error bars fall above the 5\% S.L. for all minimums considered and therefore the statistical error associated 
with $\tgH$ is enough to disallow rejection of the \pwlw hypothesis.

The biggest systematic measurement uncertainty in the CR data is the calibration of the energy. This uncertainty 
leads to an error in the reported absolute energy values of $\sim30\%$ for the AGASA \cite{AGASA} data and 
as much as $\sim$50\% for the highest energy events in the Auger data 
set. Since the probability considered here depends only on the ratio of the observed energies, it is independent of 
any constant systematic uncertainty in the energy determination. However, this probability is sensitive to energy 
errors which vary over the range considered and will thus cause uncertainty in $z^{obs}_{max}$. 

For example, if we take the maximum to be 50\% higher (but hold $\tgH=2.97$ and $N_{tot}=3567$ constant) the value of $P$ 
represented by the left most Auger point in Fig. \ref{fig:PmaxLast} changes from 1.9\% to 17\%. 
Thus the large uncertainty in $z^{obs}_{max}$ combined with the errors associated with $\tgH$ implies that 
the preliminary Auger data set does not suggest 
sufficient evidence to reject the pure \pwlw hypothesis for all events above $E_{min} = 10^{18.5}$(eV).

%=======================================================================
%=======================================================================
\section{The TP-Statistic}\label{sec:TP}
\vspace{0.05in} 

Considering the error and extra degree of freedom associated with $\tg$, an analysis of a distribution's 
adherence to the \pwlw form without reference to, or regard for, this parameter is could lead to enhanced statistical power. 
First proposed by V. Pisarenko and D. Sornette
\footnote{They studied earthquake  and financial return data.}, 
the so-called {\it TP-statistic}\cite{Pisa1,Pisa2} is a function of random variables that (in the limit of large $N$) 
tends to zero for samples drawn from a power-law, regardless of the value of $\tg$. 
(TP stands for {\it tail power}, as oppossed to TE, also introduced in \cite{Pisa1,Pisa2}, which stands for {\it tail exponential}.)
This section will describe the TP-statistic and apply it to the CR data.

The raw moments of the pdf equation (\ref{eq:pwlpdf}) are \cite{Newm}
\begin{equation}
\label{eq:pplmom}
\langle z^m \rangle_{Z} = \int_{1}^\infty z^m f_{Z}(z) dz \rightarrow
\begin{cases}
\infty & m \geq \tg \\
\frac{\tg - 1}{\tg - 1 -m}  & m < \tg.
\end{cases}
\end{equation} 
Thus power-laws with $\tg \leq 3$ have a finite mean but an infinite variance (in the limit of large N) and sample statistics 
created from these moments are not particularly helpful. However, taking the natural logarithm of $z$ allows the integrals to 
converge and one may write (for all $\tg > 1$ and $m = 0,1,2,\ldots$),
\begin{equation}
\nu_{m}  = \langle \ln^m z  \rangle_{Z} = \frac{m!}{(\tg-1)^{m}}.  \label{eq:LM}
\end{equation} 

The TP-statistic is calculated by noting that $\nu^{2}_{1} - \nu_{2}/2 = 0$. Therefore, if we use the sample analog of these 
quantities, namely 
\begin{equation}
{\hat \nu}_{m} = \frac{1}{N} \sum_{i=1}^N \ln^{m} \frac{x_{i}}{x_{min}} \label{eq:nuhat}
\end{equation}
then we can define (for all $x_{i} \geq u$),
\begin{equation}
TP(u) = \left( \frac{1}{N} \sum_{i=1}^N \ln \frac{x_{i}}{u} \right)^{2} 
              - \frac{1}{2N} \sum_{i=1}^N  \ln^{2}\frac{x_{i}}{u}. \label{eq:TP}
\end{equation}
By the law of large numbers this sample statistic tends to zero as $n\rightarrow \infty$, independent of the value of $\tg$. 
The TP-statistic allows us to test for a \pwlw like distribution without comment about the value 
of the power index. Furthermore, for any one sample we can vary $u$ from the sample minimum $X_{min}$ to the sample 
maximum $X_{max}$ and calculate the TP-statistic over the range of $x$ in the sample. 

Given complete event lists one may use equation (\ref{eq:TP}) to calculate the TP-statistic for the unbinned data. Since only the binned 
CR flux is publicly available we adapt the statistic to a binned analysis and apply it first to an example distribution with a 
cutoff and then to the CR data sets.

%=======================================================================
\subsection{An Example} \label{subsec:TP.AnExample}

\begin{figure}[h]
\begin{center}
\includegraphics[width=9.5cm, height=6cm]{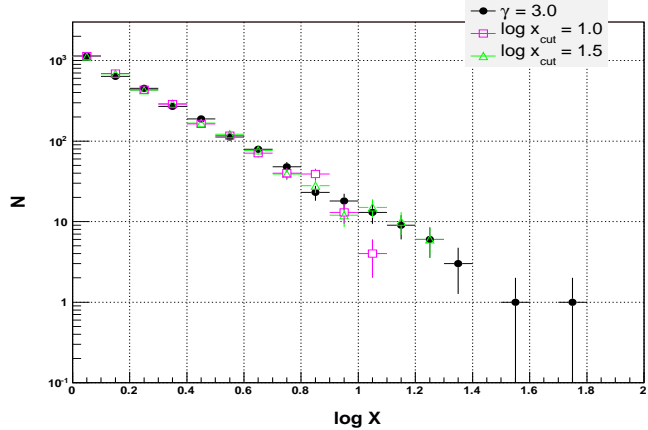} %c2FDandPPLhisto.eps
\caption{\label{fig:FDhisto} Logarithmically binned histogram of 3000 events drawn from a pure \pwlw with $\tg = 3.0$ 
and two power-laws with a cut, see equation (\ref{eq:FDpdf}). The magenta squares are drawn from the distribution 
with $\log x_{cut} = 1.0$ and the green triangles have $\log x_{cut} = 1.5$. As noted in the text, while arbitrary, 
the values of these parameters are chosen to be similar to the AGASA and Auger data (see Fig.\ref{fig:specs}).}
\end{center}
\end{figure}

In order to build intuition about the TP-statistic and its variance before studying the CR data, we first apply 
this statistic to simulated event sets drawn from both a pure \pwlw distribution and a similar distribution 
with a cut-off. The cut-off pdf is chosen so that it mimics a \pwlw for the lowest values but has an abrupt 
(and smooth) cut-off at a particular value, say $x_{cut}$. The functional form we will use here is 
\begin{equation}
g(x) = B(\tg, x_{min}, x_{cut}) \frac{x^{-\tg}}{e^{x-x_{cut}}+1}. \label{eq:FDpdf}
\end{equation}
The normalization of this pdf is $B(\tg, x_{min}, x_{cut})$, the value of which must be computed numerically.
Fig. \ref{fig:FDhisto} contains a logarithmically binned histogram of 3000 events drawn from a pure \pwlw (black circles) 
with $x_{min} = 1.0$ and $\tg = 3.0$,
and two pdf's in the from of equation (\ref{eq:FDpdf}); the magenta squares have $\log x_{cut} = 1.0$ 
and the green triangles have 
$\log x_{cut} = 1.5$. While arbitrary, the values of these parameters are chosen to be similar to the AGASA and Auger data 
(see Fig.\ref{fig:specs}).%(see Fig.\ref{fig:specs})

\begin{figure}[h]
\begin{center}
\includegraphics[width=11.1cm, height=7cm]{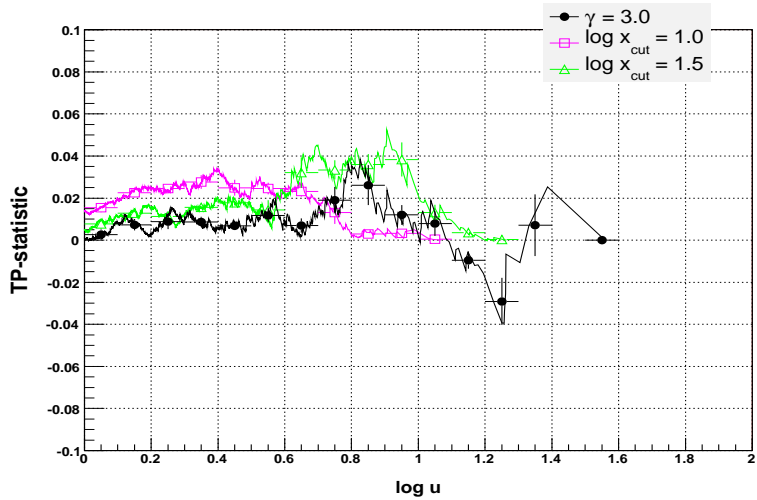} %cTP2FDandPPL.eps
\caption{\label{fig:FDTP} The TP-statistics, defined in equation equation (\ref{eq:TP}), as a function of minimum value ``$u$'' 
for the 3 sets of 3000 events plotted in Fig. \ref{fig:FDhisto}. 
Also plotted is the mean of the TP-statistic within each of the logarithmically spaced 
bins which is referred to in the text as $\otp$. The vertical error bars represent the RMS deviation of the statistic 
within each bin. Parenthetically, with increased statistics, say 10,000 events, the distinct characteristics of the 
TP-statistic for a pure power-law, a power-law with a cutoff $\log x_{cut} =1.0$ or a power-law with a cutoff $\log x_{cut} =1.5$ 
become more clearly different.}
\end{center}
\end{figure}

If we write the sorted (from least to greatest) values from a sample as $\{ X_{(1)},$$ X_{(2)}, $
$\ldots \, , X_{(N)} \}$,
the solid black line in Fig. \ref{fig:FDTP} is created by calculating $TP(u=X_{(j)})$ for each value of the 3000 events drawn 
from the pure \pwlw histogram in Fig. \ref{fig:FDhisto}. The circles represent the mean of the 
the statistic within the $i^{th}$ bin, say $\otp_{i}$, and the vertical 
error bars show the root-mean-squared deviation of the statistic within the bin. 
Note that the total number of events considered by the statistic decreases quickly from left to right which 
leads to a bias in and an increasing variance of the statistic. 

The jagged magenta line Fig. \ref{fig:FDTP} shows the most obvious deviation from the \pwlw form; it is systematically 
offset from zero for nearly all minima of the data set. Of course, with 3000 events the histograms (see Fig. \ref{fig:FDhisto}) 
are enough 
to distinguish between these two distributions. But the TP-statistic allows us to see this deviation by considering 
the entire data set (the left most magenta point in Fig. \ref{fig:FDTP}), not just by analyzing the 
events in the upper most bins. The green line in the figure shows $TP(u)$ for events drawn from equation (\ref{eq:FDpdf}) 
with $\log x_{cut} = 1.5$.
The histogram for this set is not as clearly different from the \pwlw as the magenta points and neither is the TP-statistic; 
the left-most green point shows no more deviation from zero than the power-law. However, as the minimum increases (and nears 
$x_{cut}$) the statistic moves away from zero (more noise not withstanding) and suggests that the data above the minimum 
deviate from the power-law.

It is important to note that the TP-statistic is positive for both of the cutoff distributions. 
Recall that for a pure power-law, $\nu^{2}_{1} - \nu_{2}/2 = 0$. The cutoff distribution, however, 
lacks an extended tail and will therefore have a smaller second log-moment $\nu_{2}$ as compared with 
(the square of) the first log-moment $\nu_{1}$ and will thus result in a positive TP-statistic. 
A distribution with an enhancement, rather than a cutoff, in the tail would result in a negative TP-statistic, 
since it would have a larger second log-moment (i.e. a larger ``variance''). See the Appendix ($\S$\ref{sec:App}) 
for a detailed discussion of the TP-statistic applied to the double power-law.

To quantify the significance of the TP-statistics' deviation from zero, $10^{4}$ sets of 3000 events were generated for each 
of the three distributions discussed in this section. For each set we calculate the {\it mean} TP-statistic $\otp$ within 
each of the logarithmically spaced bins. The resulting distribution of $\otp$'s within each bin is then fitted to a gaussian. 

\begin{figure}[h]
\begin{center}
\includegraphics[width=11.1cm, height=7cm]{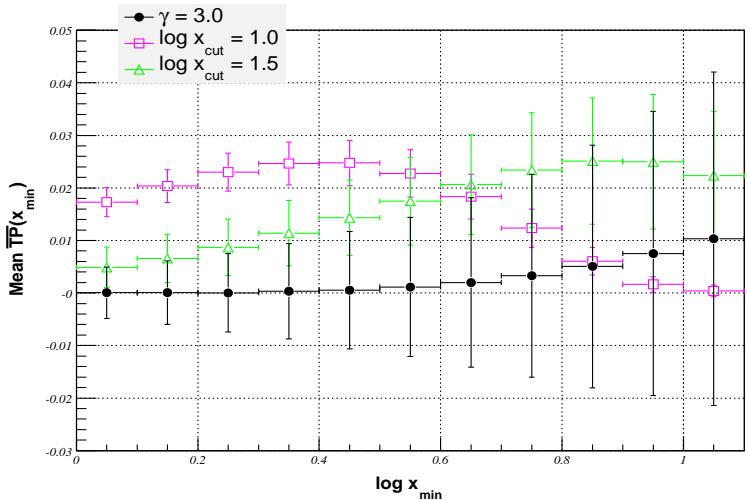}
\caption{\label{fig:FDTP_skies} The fitted mean and $1\sigma$ deviation of the $\otp$'s (see definition in text) within each bin for the 
three distributions described in the text. This plot is the result of $10^4$ simulated sets of events, where Fig.\ref{fig:FDTP} is 
one example, and where each set contains 3000 events.} 
\end{center}
\end{figure}

The black circles in Fig. \ref{fig:FDTP_skies} represent the mean of the gaussian fit to the distribution of $\otp$'s within each bin 
for a \pwlw and the error bars on the points represent the fitted $1\sigma$ deviation of the $\otp$'s. We interpret the left-most 
of these points in the following way: for 3000 events drawn from a \pwlw the ``expected value'' of $\otp$ in the 
first bin is effectively indistinguishable from zero, as expected. 

Though the statistic itself does not depend on $\tg$, the variance 
on this value does. The reason for this is that the variance of the $\otp$'s depends on the average total number of events  
greater than a given minimum, which is influenced by $\tg$. In this case the total number of events per set for minima in the 
first bin is at least a few thousand and the variance of the $\otp$'s is $\sigma_{\otp} \sim 0.005$. These errors increase from 
left to right since each successively higher bin will contain $\otp$'s based on fewer and fewer events.

The magenta squares represent the fitted mean $\otp$ as a function of $x_{min}$ for sets drawn from a \pwlw with a cut-off 
at $\log x_{cut} = 1.0$. They deviate from zero for all but the largest $x_{min}$. Furthermore, this offset is 
statistically significant for the lowest few bins of $x_{min}$, where the statistic reflects the deviation from \pwlw 
considering most of the events in the set. The green triangles show the fitted means for the $\log x_{cut} = 1.5$ distribution. They also 
display some deviation from zero, but they are not as significant since they fall near the $1\sigma$ errors for the pure 
\pwlw distribution.

\begin{figure}[h]
\begin{center}
\includegraphics[width=9.5cm, height=6cm]{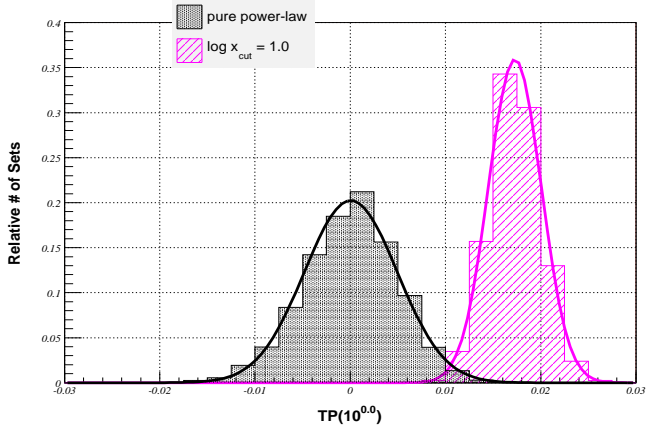}
\caption{\label{fig:FDthe1TP_100skies} The distribution of the $\otp$'s in the first bin of Fig.\ref{fig:FDTP_skies} 
(the bin with minimum $\log x_{min} = 0.0$) for 
the simulated pure \pwlw (black, shaded) and a \pwlw with a cut-off at $\log x_{cut} = 1.0$ (magenta, hatched). For these 
distributions $P_{TP} = 0.00218$ (see equation (\ref{eq:Ptp})).}
\end{center}
\end{figure}

\begin{figure}[h]
\begin{center}
\includegraphics[width=9.5cm, height=6cm]{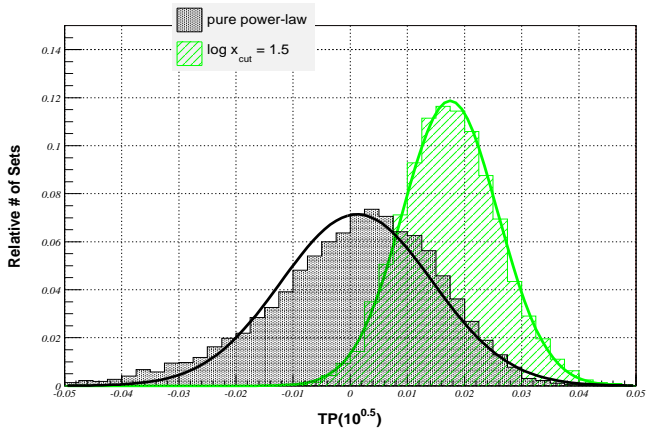}
\caption{\label{fig:FDthe2TP_100skies} The distribution of the $\otp$'s in the fifth bin of Fig.\ref{fig:FDTP_skies} 
(bin with minimum $\log x_{min} = 0.4$) for 
the simulated pure \pwlw (black, shaded) and a \pwlw with a cut-off at $\log x_{cut} = 1.5$ (green, hatched). For these 
distributions $P_{TP} = 0.298$ (see equation (\ref{eq:Ptp})).}
\end{center}
\end{figure}

Indeed, one may inquire as to which of the bins deviate the most from the simulated 
power-law. This is equivalent to asking, ``above what minimum do the data generated from this cut-off distribution maximally 
deviate from a pure power-law?'' To quantify this deviation, we use a P-value given by
\begin{equation}
P_{TP} = 1 - \frac{1}{\sqrt{2\pi}}\int_{-\beta}^{\beta} e^{-t^{2}/2}dt, \label{eq:Ptp}
\end{equation}
where 
\begin{equation}
\beta = \frac{|\mu_{1} - \mu_{2}|}{\sqrt{\sigma_{1}^{2} + \sigma_{2}^{2}}}, \label{eq:beta}
\end{equation}
$\mu_{i}$ is the mean of the fitted gaussian and $\sigma_{i}$ is the standard deviation.
We reject the pure power-law hypothesis (at the 5\% S. L.) if $P_{TP} \leq 0.05$.
The mean of the gaussian fit to the distribution of $\otp$'s for the \pwlw in the bin with minimum $\log x_{min} = 0.0$ 
is $\mu_{1} = (0.0056 \pm 5.1) \times 10^{-3}$ with a standard deviation $\sigma_{1} = 4.9 \times 10^{-3}$. 
The mean of the fitted gaussian for the $\log x_{cut} = 1.0$ distribution in this 
bin is $\mu_{2} = (1.7 \pm 0.28) \times 10^{-2}$ with a standard deviation $\sigma_{2} = 2.8 \times 10^{-3}$.
Therefore, the significance level of the deviation is $P_{TP} = 2.18 \times 10^{-3}$ and we can reject the pure power-law 
hypothesis for this distribution. The 
distribution of the $\otp$'s for this bin is plotted in Fig. \ref{fig:FDthe1TP_100skies} for the pure \pwlw (black, shaded) 
and the $\log x_{cut} = 1.0$ (magenta, hatched) pdf. The maximum deviation for the $\log x_{cut} = 1.5$ pdf 
occurs in the bin with minimum $\log x_{min} = 0.4$ and the corresponding distributions of $\otp$ are 
plotted in Fig. \ref{fig:FDthe2TP_100skies}. The significance of this deviation is lower; $P_{TP} = 0.298$.

%=======================================================================
\subsection{The Cosmic Ray Data} \label{subsec:TP.Data}

In order to apply the TP-statistic to the CR data, Monte-Carlo simulations were conducted and analyzed in a manner similar to 
that discussed in $\S$\ref{subsec:TP.AnExample}; we generate $10^{4}$ sets of events from the reported flux and the resulting 
distribution of $\otp$ (within each bin) is fitted to a gaussian. Since the significance of deviation from zero 
depends on both the power index and the number of events, we will compare each of the Auger and AGASA data sets with a unique 
power-law. We will take the AGASA experiment to have $1916$ events above $\log E_{min} = 18.8$ and we will compare the 
resulting TP-statistics with those of a \pwlw with the same minimum and $\tg = 2.80$. The Auger spectrum 
has a power-index estimate of $2.97$ considering all of the data above $\log E_{min} = 18.5$ and a total of 3570 events, 
so we will therefore 
compare the TP-statistics arising from the Auger flux to those of a pure \pwlw with these parameters.

\begin{figure}[h]
\begin{center}
\includegraphics[width=11.1cm, height=7cm]{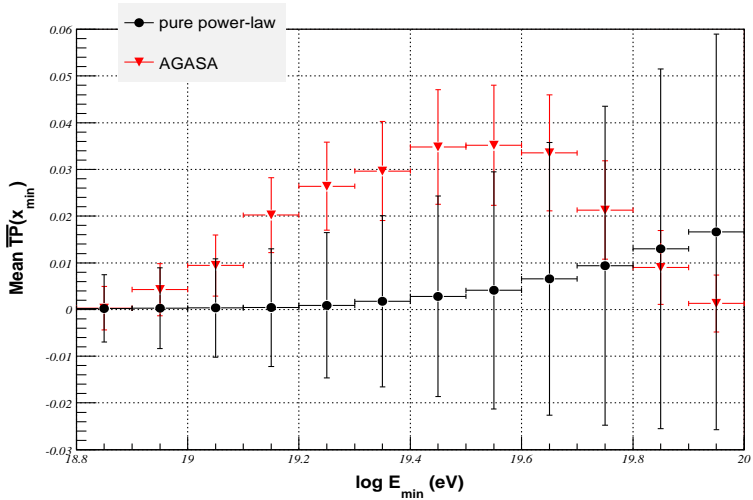} 
\caption{\label{fig:TPstatsAGA} The fitted mean and $1\sigma$ deviation of the $\otp$'s (see definition on the text) within 
each bin for the AGASA spectrum (red triangles) and a pure \pwlw distribution (black circles). This plot is the result 
of $10^4$ simulated sets of events where each set contains 1916 events and the power-law has index $\tg = \tgH_{AGASA}$.}
\end{center}
\end{figure}

\begin{figure}[h]
\begin{center}
\includegraphics[width=9.5cm, height=6cm]{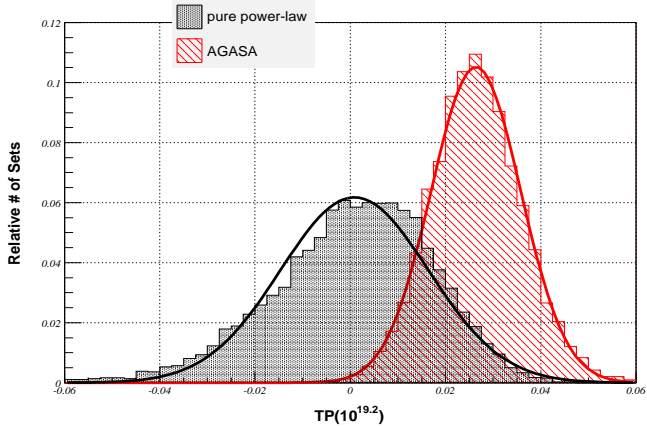} 
\caption{\label{fig:TheTPAGA} The distribution of $\otp$'s in the fifth bin of Fig.\ref{fig:TPstatsAGA}
(the bin with minimum $E_{min} = 10^{19.2}$(eV)) for the pure power-law (black, shaded) and the 
AGASA spectrum (red, hatched). For these distribution $P_{TP} = 0.161$ (see equation (\ref{eq:Ptp})).}
\end{center}
\end{figure}

The application of this scheme to the AGASA spectrum is plotted in Fig. \ref{fig:TPstatsAGA} in red triangles. The 
black circles represent average TP-statistic value for data drawn from a pure \pwlw with $\tgH_{AGASA}$. 
Both plots have $N=886$ events per sky. The error bars on each point represent the 1-sigma deviation of 
the gaussian fit to the distribution of the mean TP-statistic. Since the AGASA values do not significantly deviate from 
zero (or the \pwlw values) this plot suggests that the AGASA distribution does not significantly deviate from a pure power-law. 
The most significant deviation occurs in the bin with minimum $10^{19.2}$(eV) and gives $P_{TP} = 0.161$, which is 
consistent with the P-value for this bin discussed in $\S$\ref{sec:DLV}. 
These distributions are plotted in Fig \ref{fig:TheTPAGA}.

\begin{figure}[h]
\begin{center}
\includegraphics[width=11.1cm, height=7cm]{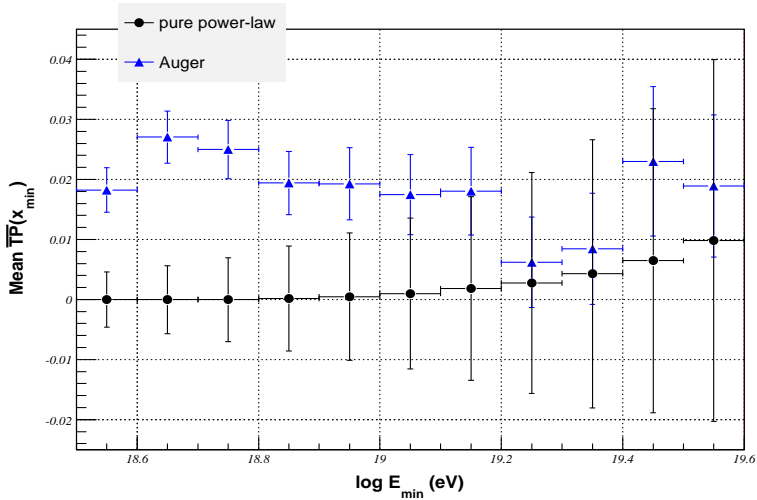}
\caption{\label{fig:TPstatsAUG} The fitted mean and $1\sigma$ deviation of the $\otp$'s (see definition on the text) within 
each bin for the Auger spectrum (blue triangles) and a pure \pwlw distribution (black circles). This plot is the result 
of $10^4$ simulated sets of events where each set contains 3570 events and the power-law has index $\tg = \tgH_{Auger}$.}
\end{center}
\end{figure}

\begin{figure}[h]
\begin{center}
\includegraphics[width=9.5cm, height=6cm]{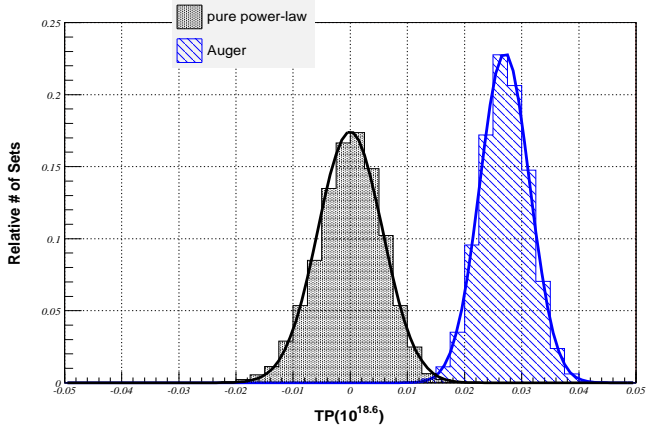}
\caption{\label{fig:TheTPAUG}  The distribution of $\otp$'s in the second bin of Fig.\ref{fig:TPstatsAGA}
(the bin with minimum $E_{min} = 10^{18.6}$(eV)) for the pure power-law (black, shaded) and the 
Auger spectrum (blue, hatched). For these distribution $P_{TP} = 1.54 \times 10^{-4}$ (see equation (\ref{eq:Ptp})).}
\end{center}
\end{figure}

The simulation results from the Auger spectrum are plotted in Fig. \ref{fig:TPstatsAUG}.
This plot shows deviation from a \pwlw for the lowest minimums considered. 
For the bin with minimum $\log E_{min} = 18.6$ we find $P_{TP} =  1.54 \times 10^{-4}$.
Thus we can say that the Auger spectrum with energies greater than $10^{18.6}$(eV) deviate from a 
\pwlw by $\sim 3.78\sigma$, where $\sigma^2 = \sigma_{1}^{2} + \sigma_{2}^{2}$.
The distribution of $\otp$'s for this minimum energy is plotted in Fig. \ref{fig:TheTPAUG}.

Since the TP-statistic nearly eliminates the need to estimate $\tg$, the biggest systematic uncertainty in analyzing the CR data 
with the TP-statistic is likely to be errors in the event energies. 
Similar to the $P$-value discussed in $\S$\ref{sec:DLV}, it is only the relative energy errors which can effect the result, since 
the TP-statistic depends only on the ratio. However, any elongation of the observed spectrum brought about by this relative uncertainty 
effect the TP-statistic. Without further study of the CR energy systematics, we cannot draw a conclusion from the 
$\sim3.78\sigma$ deviation in Fig. \ref{fig:TheTPAUG}.

%=======================================================================
\vspace{0.2in}
\section{Summary}
\vspace{0.15in}

In $\S$\ref{sec:data} we use the reported (AGASA and Auger) CR fluxes to discuss the \pwlw form and illustrate 
the logarithmically binned estimates of the power index $\tg$. 
The probability $P$ that the maximum value of a sample drawn from a \pwlw is less than or equal to a particular value 
is defined in equation (\ref{eq:Pxc}). 
Using reasonable estimates for $\tg$,  $N_{tot}$ and $z^{obs}_{max}$ from the CR data sets we calculate $P$ in $\S$\ref{sec:DLV}.
The value of $P$ is used to test the null hypothesis that these data sets follow a \pwlw. The AGASA data give no reason 
to reject the hypothesis; $P_{AGASA} \sim 8.4\%$ for the data with $\log E(eV) \geq 18.8$. 
The Auger data give more reason to reject the null hypothesis, $P_{Auger} \sim 1.9\%$ for the data with $\log E(eV) \geq 18.5$. 
However, consideration of the errors on $\tgH$ prevent any solid conclusion. 

For the purpose of statistical analysis it would be useful to eliminate, or at least minimize, the importance of $\tg$. 
The TP-statistic tends (asymptotically) to zero regardless of the value of $\tg$ and is the subject of $\S$\ref{sec:TP}.
We apply the TP-statistic to the CR data sets using a Monte-Carlo method described in $\S$\ref{subsec:TP.Data}. 
The AGASA data give a value of $P_{TP} = 0.161$ for energies greater than $10^{19.2}$(eV). 
a value consistent with the $P$-value discussed in $\S$\ref{sec:DLV} (Fig.\ref{fig:PmaxLast}).
The preliminary Auger flux 
results in a TP-statistic with more significant deviation from the power-law form: $P_{TP} = 1.54 \times 10^{-4}$ for  
$E_{min} = 10^{18.6}$(eV). 
Comparing this value with the $P$-value for this bin derived in $\S$\ref{sec:DLV}, namely $P \sim 2 \times 10^{-2}$, 
illustrates the power of the method based on the TP-statistic which is essentially independent of gamma.
Better understanding of the relative errors on the CR energies should lead to a definitive conclusion on the
question of a cut-off in the CR spectrum.

%\newpage 

\newpage 
%\include{Appendix}

%=======================================================================
\vspace{0.2in}
\section{Appendix} \label{sec:App}
\vspace{0.15in}

In $\S$\ref{subsec:TP.AnExample} we state that the TP-statistic will be distinctly positive for distributions 
which contain a tail-suppression and negative for distributions which contain a tail-enhancement (relative 
to the pure \pwlw form). In this section we numerically compute the TP-statistic for a ``double power-law'' 
distribution and describe the parameter space associated with this statistic.

Consider the following probability distribution function:
\begin{equation}
f(x) = \label{eq:pdfBend}
\begin{cases}
A(x_{min}, x_{bend}, \tg, \delta) x^{-\tg} &  x_{min} \leq x < x_{bend} \\
B(x_{min}, x_{bend}, \tg, \delta) x^{-\delta} &  x_{bend} \leq x < \infty,
\end{cases}
\end{equation}
where $A(x_{min}, x_{bend}, \tg, \delta)$ and $B(x_{min}, x_{bend}, \tg, \delta)$ are chosen such that 
\begin{equation}
\lim_{x \rightarrow x^{+}_{bend}} f(x) = \lim_{x \rightarrow x^{-}_{bend}} f(x) \nonumber
\end{equation}
and 
\begin{equation}
\int_{x_{min}}^{\infty} f(x)dx=1.\nonumber
\end{equation}
This distribution follows a power-law with index $\tg$ for $x_{min} \leq x < x_{bend}$, and $\delta$ for $x \geq x_{bend}$.\\ 

Given the parameter set $\{x_{min}, x_{bend}, \tg, \delta\}$, we define the 
TP-statistic for this distribution as
\begin{equation}
TP(u) = \left[ \int_{u}^{\infty} \ln \left( \frac{x}{u} \right) f(x)dx \right]^{2} - 
        \frac{1}{2} \int_{u}^{\infty} \ln^{2} \left( \frac{x}{u} \right) f(x)dx. \label{eq:TPbend}
\end{equation}
For $u \geq x_{bend}$ and/or $\tg = \delta$ equation (\ref{eq:TPbend}) is identically zero since it is equal 
to $\nu_1^2 - 1/2\nu_2$ (see equation (\ref{eq:LM})). However, equation (\ref{eq:TPbend}) is non-trivial 
when $x_{min} \leq u < x_{bend}$ and $\tg \neq \delta$. 
In what follows, 
we calculate $TP(u)$ for $x_{min} \leq u < x_{bend}$ and various values of $x_{bend}$ and $\delta$ with $x_{min} = 1$ and $\tg = 3$ fixed.

\begin{figure}[h]
\begin{center}
\includegraphics[width=9.5cm, height=6cm]{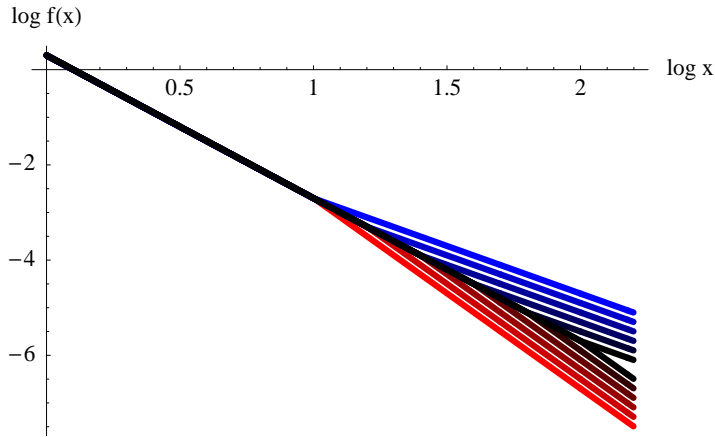}
\caption{\label{fig:appXbendPDF}  A plot of $\log f(x)$ (see equation (\ref{eq:pdfBend})) versus $\log x$ 
with $\delta = \tg \pm 1$ for several choices of $\log x_{bend}$ (namely, for $\log x_{bend}$ varying from 1 to 2 in steps of 0.2). 
The red curves correspond to $\tg<\delta = 4$ (tail-suppression) and the blue curves have $\tg>\delta = 2$ 
(tail-enhancement).  The more black the color of the curve, the larger $\log x_{bend}$.}
\end{center}
\end{figure}

\begin{figure}[h]
\begin{center}
\includegraphics[width=9.5cm, height=6cm]{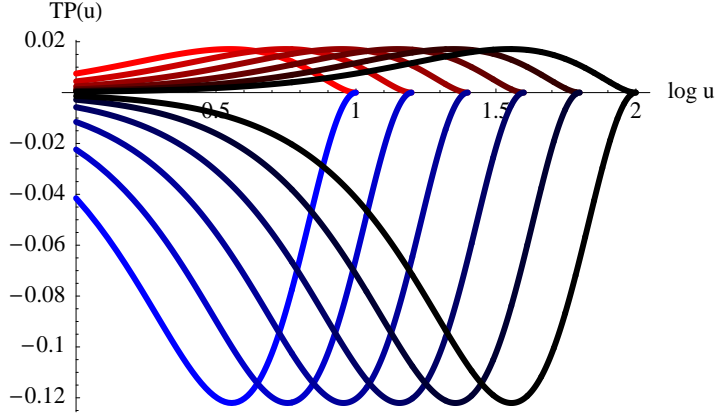}
\caption{\label{fig:appXbendTP}  A plot of $TP(u)$ (see equation (\ref{eq:TPbend})) for each of the distributions 
plotted in Fig.\ref{fig:appXbendPDF}. Those distributions with tail-suppression (red) have $TP(u)>0$ and those 
with tail-enhancement (blue) have $TP(u)<0$.}
\end{center}
\end{figure}

Fig.\ref{fig:appXbendPDF} contains a plot of $\log f(x)$ versus $\log x$ with $\delta = \tg \pm 1$ for several choices of $\log x_{bend}$ (namely, for $\log x_{bend}$ varying from 1 to 2 in steps of 0.2). 
The red curves correspond to $\tg<\delta = 4$ (tail-suppression) and the blue curves have $\tg>\delta = 2$ 
(tail-enhancement).
The TP-statistic for each of these distributions is shown in Fig.\ref{fig:appXbendTP} as a function of $u$. 
Examination of Fig.\ref{fig:appXbendTP} suggests the following conclusions for a given $\tg$ and $\delta$: 
\begin{itemize}
\item $TP(u)$ is positive for all values of $u$ and $x_{bend}$ if and only if $\tg<\delta$, and it is negative if and 
only if $\tg>\delta$.
\item For $x_{bend}$ much greater than $x_{min}$, $TP(u=x_{min})$ is approximately zero. Specifically, 
as $x_{bend}/x_{min} \rightarrow \infty$, $TP(x_{min}) \rightarrow 0$.
\item The location of the maximum deviation, say $u_0$ where 
\begin{equation}
\left( \frac{\partial}{\partial u} TP(u) \right)_{u= u_0} = 0, \label{eq:U0def}
\end{equation}
is highly correlated with the location of the bend $x_{bend}$. Indeed, we have found that 
there is a linear relationship between $\log u_0$ and $\log x_{bend}$ and that this relationship is independent of whether 
$\tg$ is less than or greater than $\delta$.
\item The maximum deviation of the TP-statistic, i.e. $TP(u_0)$, is independent of $\log x_{bend}$.
\end{itemize}

\begin{figure}[h]
\begin{center}
\includegraphics[width=9.5cm, height=6cm]{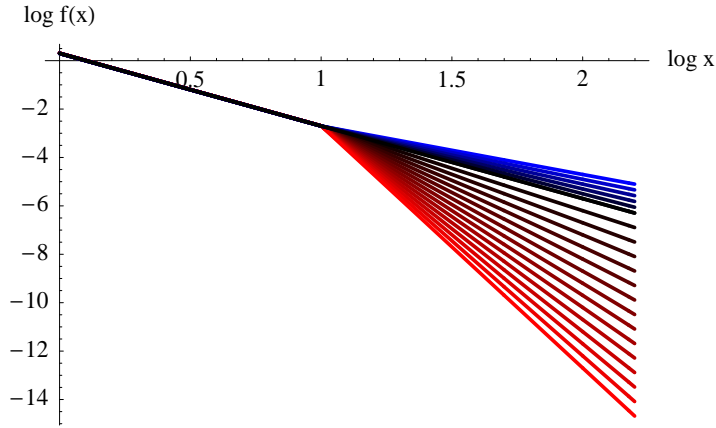}
\caption{\label{fig:appDeltaPDF}  A plot of $\log f(x)$ (see equation (\ref{eq:pdfBend})) versus $\log x$ 
with $\log x_{bend} = 1.0$ and $\tg =3$. The blue curves have $2 \leq \delta < 3$ (i.e. $\delta-\tg<0$) and the 
red curves have $3 < \delta < 10$ (i.e. $\delta-\tg>0$). The more black the color of these curves, the closer 
$\delta$ is to $\tg$.}
\end{center}
\end{figure}

To isolate the effects of power index choice, consider the family of distributions where $\log x_{bend} = 1.0$ 
is fixed but $\delta$ is allowed to vary. Since the integrals in equation (\ref{eq:TPbend}) only converge if 
$\delta \geq 2$, the minimum $\delta$ we can choose is $\delta = 2$. There is no upper bound on $\delta$ so 
we vary this parameter over the interval $2 \leq \delta < 3$ in steps of 0.2 and over the interval 
$3 < \delta < 10$ in steps of 0.5. Fig.\ref{fig:appDeltaPDF} contains a plot of $\log f(x)$ versus $\log x$ 
with $\log x_{bend} = 1.0$ and $\tg =3$. The blue curves have $2 \leq \delta < 3$ (i.e. $\delta-\tg<0$) and the 
red curves have $3 < \delta < 10$ (i.e. $\delta-\tg>0$). The more black the color of these curves, the closer 
$\delta$ is to $\tg$.

\begin{figure}[h]
\begin{center}
\includegraphics[width=9.5cm, height=6cm]{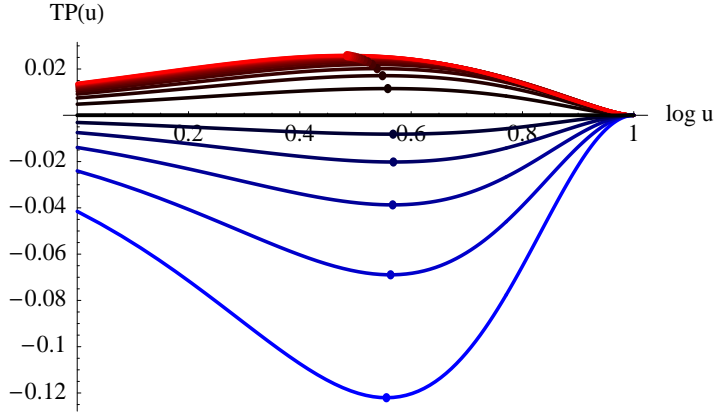}
\caption{\label{fig:appDeltaTP}  A plot of $TP(u)$ (see equation (\ref{eq:TPbend})) for the distributions 
plotted in Fig.\ref{fig:appDeltaPDF}. 
The colored points on these curves show where each curve maximally deviates from zero; the coordinates of these 
points are $\{ u_0, TP(u_0) \}$ for each curve (see equation (\ref{eq:U0def})).}
\end{center}
\end{figure}

Fig.\ref{fig:appDeltaTP} contains a plot of $TP(u)$ for the distributions plotted in Fig.\ref{fig:appDeltaPDF}. 
As noted earlier, $TP(u)>0$ if and only if $\delta - \tg>0$ and $TP(u)<0$ if and only if $\delta - \tg<0$.
The colored points on these curves show where each curve maximally deviates from zero; the coordinates of these 
points are $\{ u_0, TP(u_0) \}$ for each curve (see equation (\ref{eq:U0def})). These points show a weak 
dependence of $\log u_0$ on $\delta$, for a given $\log x_{bend}$.

\begin{figure}[h]
\begin{center}
\includegraphics[width=9.5cm, height=6cm]{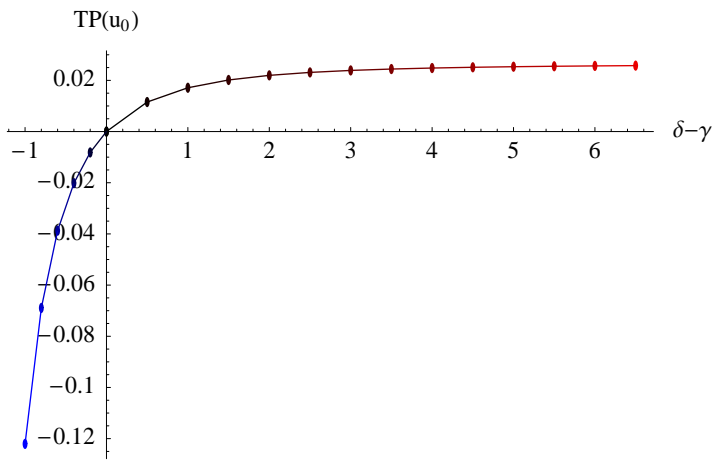}
\caption{\label{fig:appDeltaTPu0}  A plot of $TP(u_0)$ (see equations (\ref{eq:pdfBend}) and (\ref{eq:U0def})) 
versus $\delta-\tg$ for each of the points in 
Fig.\ref{fig:appDeltaTP}. Note that $TP(u_0) \approx 0.025$ for $\delta-\tg \gg 1$.}
\end{center}
\end{figure}

The value of the maximum deviation $TP(u_0)$ also shows dependence on $\delta$. 
In Fig.\ref{fig:appDeltaTPu0} we plot $TP(u_0)$ versus $\delta-\tg$ for each of the points in 
Fig.\ref{fig:appDeltaTP}. These plots suggest the following: 
\begin{itemize}
\item For $-1 \leq \delta-\tg \lesssim 1$ (blue and black), a small change in $\delta$ will lead to a large change in $TP(u_0)$. 
\item If $\delta-\tg \gg 1$ (bright red), however, a large change in $\delta$ will result in a small change in $TP(u_0)$. 
This case is of particular interest since a large $\delta$ will mimic the cutoff distribution defined 
in equation (\ref{eq:FDpdf}). 
\item By inspection of Fig.\ref{fig:appDeltaTPu0} we note that $TP(u_0) \approx 0.025$ for $\delta-\tg \gg 1$. 
\item Comparison with Fig.\ref{fig:appXbendTP} suggests that the limiting value of $TP(u_0)$ is 
roughly independent of $x_{bend}$.
\end{itemize}

The studies described in this section show that the TP-statistic can distinguish tail-suppressed ($\delta-\tg > 0$)
from tail-enhanced ($\delta-\tg < 0$) distributions, i.e. $TP(u)>0$ if and only if $\delta - \tg>0$ and $TP(u)<0$ 
if and only if $\delta - \tg<0$. 
Furthermore, they show that in the limiting case of $\delta-\tg \gg 1$ the most important parameter in determining $u_0$ 
is $x_{bend}$ but that the limiting value of $TP(u_0)$ is roughly independent of $x_{bend}$ and $\delta - \tg$.


\begin{thebibliography}{99}
\bibitem{Newm} M. E. J. Newman, Contemporary Physics {\bf 46}, 323-351 (2005).
\bibitem{Claus} A. Clauset, M. Young and K.S. Gleditsch, {\bf arXiv:physics/0606007} (2006).
\bibitem{G} K. Greisen, Phys. Rev, Lett. {\bf 16}, 748 (1966).
\bibitem{ZK} G.T. Zatsepin and V.A. Kuzmin, JETP. Lett. 4 {\bf 78}, (1966).
\bibitem{Bahc} John N. Bachall and Eli Waxman {\bf arXiv:hep-ph/0206217 v5} (2003).
\bibitem{Auger} Auger Collaboration, Proceedings $29^{th}$ ICRC, Pune, India, {\bf 10} 115 (2005), {\bf arXiv:astro-ph/0604114}.
\bibitem{Yama} T. Yamamoto and The Pierre Auger Collaboration, {\bf arXiv:astro-ph/0601035 v1} (2006).
\bibitem{AGASA} M. Takeda {\it et al.}, Phys. Rev. Lett. {\bf 81} 6 (1998).
\bibitem{maxlik} Unbinned maximum likelihood methods have less error and bias when applied to power-law (or similar) distributions 
than binned methods\cite{Golds}. They can also be modified to include energy error and variable acceptance information\cite{Howe}. 
Lacking this information, we use the logarithmically binned estimate of $\tg$ where necessary. The minimum variance for 
{\it any} estimator of $\tg$ is given by the Cramer-Rao lower bound; $\sigma_{\hat{\tg}} \geq \left( \frac{\tg+1}{N} \right)^{1/2}$.
\bibitem{Golds} M. L. Goldstein, S. A. Morris, and G. G. Yen, Eur. Phys. J. B. {\bf 41}, 255-258 (2004).
\bibitem{Howe} L.W. Howell, {\it Statistical Properties of Maximum Likelihood Estimators of Power Law Spectra Information}, 
NASA/TP-2002-212020/REV1, Marshall Space Flight Center, Dec., 2002.
\bibitem{Pisa1} V. Pisarenko, D. Sornette and M Rodkin, 
{\it Deviations of the Distributions of Seiesmic Energies from the Gutenberg-Richter Law}, Computational Seismology {\bf 35}, 
138-159 (2004), {\bf arXiv:physics/0312020}.
\bibitem{Pisa2} V. Pisarenko, D. Sornette, {\it New Statistic for Financial Return Distributions: \pwlw or exponential?}, 
{\bf arXiv:physics/0403075}.
\end{thebibliography}
\end{document}